\documentclass[a4paper,fleqn,usenatbib]{mnras}

\usepackage{newtxtext,newtxmath}

\usepackage[T1]{fontenc}
\usepackage{ae,aecompl}

\usepackage{graphicx}	
\usepackage{amsmath}	

\usepackage[table]{xcolor}

\usepackage{caption}
\DeclareCaptionType{equ}[Expression][]

\usepackage{array}

\title[The Locus Algorithm Quasar Pointings Catalogue]{A Catalogue of Locus Algorithm Pointings for Optimal Differential Photometry for 23,779 Quasars}

\author[O. Creaner et al.]{
Ois\'{i}n Creaner,$^{1,2,5}$\thanks{E-mail: creanero@gmail.com, oocreaner@lbl.gov(OC)}
Kevin Nolan,$^{1}$
David Grennan,$^{4}$
Niall Smith$^{3}$
and Eugene Hickey$^{1}$
\\
$^{1}$Technological University Dublin, Tallaght Campus, Dublin 24, Ireland\\
$^{2}$Dublin Institute for Advanced Studies, 31 Fitzwilliam Place, Dublin 2, Ireland\\
$^{3}$Cork Institute of Technology, Cork City, Ireland\\
$^{4}$Raheny Observatory, Raheny, Dublin 5, Ireland\\
$^{5}$Lawrence Berkeley National Laboratory, 1 Cyclotron Road, Berkeley, California, USA
}

\date{Accepted XXX. Received YYY; in original form ZZZ}

\pubyear{2020}

\begin{document}
\label{firstpage}
\pagerange{\pageref{firstpage}--\pageref{lastpage}}
\maketitle

\begin{abstract}
This paper presents a catalogue of optimised pointings for differential photometry of 23,779 quasars extracted from the Sloan Digital Sky Survey (SDSS) Catalogue and a score for each indicating the quality of the Field of View (FoV) associated with that pointing.
Observation of millimagnitude variability on a timescale of minutes typically requires differential observations with reference to an ensemble of reference stars.  
For optimal performance, these reference stars should have similar colour and magnitude to the target quasar.  In addition, the greatest quantity and quality of suitable reference stars may be found by using a telescope pointing which offsets the target object from the centre of the field of view.
By comparing each quasar with the stars which appear close to it on the sky in the SDSS Catalogue, an optimum pointing can be calculated, and a figure of merit, referred to as the ``score'' calculated for that pointing.  
Highly flexible software has been developed to enable this process to be automated and implemented in a distributed computing paradigm, which enables the creation of catalogues of pointings given a set of input targets.
Applying this technique to a sample of 40,000 targets from the 4th SDSS quasar catalogue resulted in the production of pointings and scores for 23,779 quasars.  This catalogue is a useful resource for observers planning differential photometry studies and surveys of quasars to select those which have many suitable celestial neighbours for differential photometry
\end{abstract}




\section{Introduction}
\label{Introduction}

The variation of apparent brightness of all but the most extremely variable astronomical objects is dominated by atmospheric effects \citep{young1991precise}. According to \citet{giltinan2011using} we can assume that for narrow angular separations, the atmospheric effects are correlated. \citet{burdanov2014astrokit} characterised this assumption as being most correct when target and references are within a radius of 5-7 arcminutes of one another, but remains practical out to at least 15 arcminutes. This assumption underpins the technique of differential photometry, whereby a series of observations are made of the target and one or more references, and the difference between the target and the reference(s) is used to plot a light-curve \citep{milone2011high}.

Current practice in differential photometry is to use a collection of reference stars known as an ensemble, which allows for intrinsic variability of any one reference contributes little to the variation of the ensemble, and whereby each candidate reference can be compared against the ensemble, and should they prove variable, they are removed from it \citep{everett2001technique, honeycutt1992ccd}. A greater number of references can therefore be seen to permit greater precision and flexibility in excluding previously unknown variables among the reference stars.

High-precision photometry demands that the references used to compare against the target should be similar in magnitude and colour to the target object \citep{young1991precise, milone2011high, budding2007introduction}. The reasons for a close magnitude match are to avoid detector saturation (for bright references) and to avoid low signal-to-noise ratios (for faint references)\citep{young1991precise, milone2011high}. The rationale for closely matching colour indices is based on second-order extinction factors which are wavelength-dependent \citep{milone2011high}.  \citet{young1991precise} suggests that best practice is to use reference stars which are within a maximum of
$\pm$ 0.3 mag of the target in Johnson B-V colour index \citep{young1991precise}.

This work feeds into observational studies producing lightcurves of quasar luminosities. Previous work, for example \citet{2009MNRAS.397..558K} demonstrate lightcurve analyses such as detrending techniques.

Time-resolved precision photometry has the potential to infer very small scale structures in astrophysical jets at a scale which is not possible with direct imaging (Smith et al. 2008).
Despite the knowledge that reference stars which are similar to the source provide the best approach, there has been no systematic attempt to define the optimum FoV
around a target source in which parameters such as colour, magnitude, field-crowding and field orientation have been used to determine the optimum pointings.

\begin{figure}
        \center{\includegraphics[width=0.47\textwidth]
        {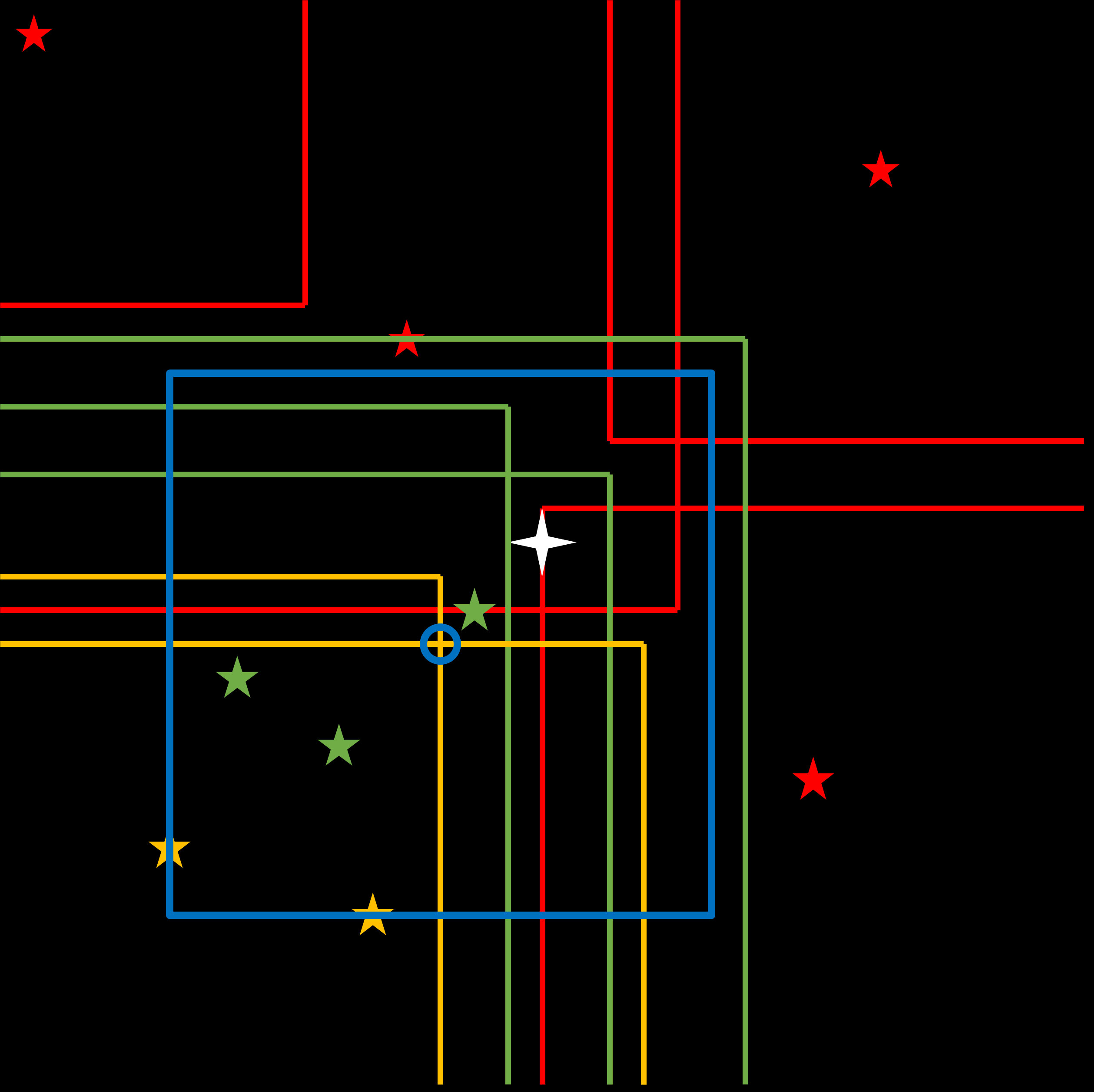}}
        \caption{\label{fig:locus_figure} A worked example of the Locus Algorithm.  The target (white) has been included in a Field of View (blue square) with five reference stars (yellow and green) by offsetting the centre of the telescope pointing (blue circle) to a point defined by the intersection of the loci of points upon which the centre of the FoV may be centred and still include the given reference (coloured lines).  Copied from \citet{creaner2016thesis}.  A more detailed treatment of this example is given in \citet{locuspaper}. }
      \end{figure}

In \citet{locuspaper} we describe an algorithm, the Locus Algorithm, which identifies the pointing for the which the resultant observational FoV includes the target and the most photometrically appropriate reference stars available. When provided with a list of targets, the algorithm calculates a score for each, allowing the user to identify targets with large numbers of high-quality reference stars \citep{creaner2016thesis, locuspaper, creaner2010large}. Figure \ref{fig:locus_figure} shows a worked example of the Locus Algorithm in operation. This paper presents a catalogue of pointings constructed using the Locus Algorithm technique. 

For any target whose field is being optimised for differential photometry, the Locus Algorithm requires as input a set of parameters associated with the observing system, the target and other stars in the vicinity of the target.  The parameters for the observing system are: (1) the size of the field of view, (2) the resolution of the instrument for which the outputs are to be optimised, (3) the maximum permitted difference between the target and any given reference in magnitude ($\Delta{}m_{max}$) and (4) the maximum permitted difference in colour index between the target and the references ($\Delta{}col_{max}$).  The catalogue produced is based on a constant frame size using an equatorial mount aligned along a North-South East-West axis.

In the implementation demonstrated here, which uses SDSS \citep{abazajian2009seventh} as the data source, the SDSS \textit{ugriz} filter for which the system is optimised is also provided as a parameter.  For the target and the other stars in the vicinity, the position (RA, Dec) and magnitudes (e.g. SDSS \textit{ugriz}) are required.  For details of its operation, see \citet{creaner2016thesis} and \citet{locuspaper}.

The technique used to generate this catalogue described as in \citet{locuspaper} can be used to generate similar optimised catalogues from other sky surveys so long as they contain RA, Dec and Magnitude information on a per-object basis. This paper describes the application of the Locus Algorithm to identifying optimum pointing for the differential photometry of quasars from SDSS Quasar Catalogue 4 \citep{schneider2007sloan}. This algorithm has been applied to a collection of targets in the form of 40,000 quasars in the SDSS footprint and a set of pointings and scores have been generated to enable optimised photometry for 23,779 of those quasars \citep{creaner2016thesis}.

\section{Inputs for an SDSS quasar pointings catalogue}
\label{Parameters}
As indicated in Section \ref{Introduction}, any given catalogue of targets to be used for optimised differential photometry using The Locus Algorithm requires a set of parameters, some of which are telescope specific.  We present a first catalogue of quasar targets with optimized pointings, using the SDSS as source material.  Field of View and resolution limit parameters typical of the class of telescopes to be used for such studies were chosen, and provide for the possibility of generating a catalogue usable ``out of the box'' on a large number of typical telescopes. These parameters are discussed below.

\subsection{Parameters of targets and references}
\label{Targets}

\textbf{Target list:} The target list for this catalogue was derived from the 4th SDSS quasar catalogue \citep{schneider2007sloan}. Of the 77,429 quasars in that catalogue, 40,000 were processed for technical reasons. These were batched into groups of 1,000.  Each batch of 1,000 constituted a grid job and was processed using a \textit{glite}-based grid computing system \citep{glite} described in detail in \citet{creaner2016thesis} and \citet{locus_software_paper}.

\textbf{Reference star data:}  Data regarding the stars in the sky surrounding each target was identified by SQL queries to the SDSS Catalogue Archive Server \citep{SDSSCAS} and accessed via a copy of the Object List files (FITS format files containing lists of detected objects) \citep{SDSSDAS} from SDSS which had been loaded into a grid network storage system known as the Local File Catalogue  (LFC) as shown in \citet{locus_software_paper}.

\subsection{Parameters of Observing Conditions}
\label{Observing}

The following observational parameters were selected based on ideal conditions for observation at high-quality sites for instruments similar to that at Blackrock Castle Observatory (BCO).  These conditions are parameterised in the software described in \citet{creaner2016thesis} and \citet{locus_software_paper} and this can be set to other suitable values for subsequent catalogues or individual analyses of a given target.

\textbf{Field of view size:} The Field of View selected for this catalogue was a 10 arcminute square, typical of many narrow-field telescopes.  

\textbf{Resolution:} The Resolution parameter used in this catalogue was 1 arcsecond.  While this is optimistic for non-adaptive optics instruments, this was chosen to provide an example of the output for a telescope with optimum seeing.  For non-crowded fields, this parameter is not a dominant contributor to candidate selection criteria.

\textbf{Magnitude difference limit:} $\Delta{}m_{max} = \pm 2.0$ magnitudes was selected, as it meant that the brightest reference was $\sim$40 times brighter than the faintest, well within the dynamic range for common detectors.

\textbf{Colour difference limit:} $\Delta{}col_{max} = \pm 0.1$  magnitudes was selected.  This is well inside the 
$\pm$
0.3 limit suggested by \citet{young1991precise} By providing a more strict limit, this can provide a baseline from which subsequent observations can expand.

\textbf{Filter:} This catalogue is based on observations in SDSS's central \textit{r} band.  As shown in \citet{locuspaper}, this means that quasars are compared in terms of their SDSS \textit{g}, \textit{r}, and \textit{i} magnitudes.

\subsection{Scoring}

The current implementation of the scoring system is designed to match the colour of the target with the colour of the references as closely as is possible.  References with a more similar colour to the target are given a higher Rating than those which are more dissimilar as shown in Expression \ref{Rating}.  The sum of all Ratings of candidate reference stars centred on a given pointing is the score for that pointing.

\begin{equ}
  \begin{equation}
  \begin{split}
  R &=\left( 1- \left|{\frac{\Delta g - \Delta r}{\Delta col_{max}}}\right| \right) \times \left(1-\left|{\frac{\Delta r - \Delta i}{\Delta col_{max}}}\right|\right) \\
  S&=\sum_{ref}^{FoV}R 
  \end{split}
    \end{equation}
\caption{\label{Rating}Definition of the rating (\textit{R}) of a reference star.  \textit{g}, \textit{r} and \textit{i} are SDSS magnitudes. ${\Delta}$ represents a difference between the target and a given reference star. ${\Delta col_{max}}$ is the maximum permitted colour difference.  Score (\textit{S}) is defined as the sum of the Ratings of all references in the FoV.  Mathematically simplified version of the rating and scoring system defined in \citep{locuspaper}. }
\end{equ}


\section{Results}
\label{Results}
The parameters above were submitted to the software system described in \citet{locus_software_paper}, the code for which is available on GitHub at \citet{githubrepo}.  The results were output to the LFC on Grid Ireland using the system described in \citet{grid_system_paper} and subsequently were collated into the catalogue, which is available on Zenodo at \citet{ZenodoQuasarCatalogue}.  The performance metrics of the grid computing system as used to create this catalogue are  described in \citet{grid_metrics_paper}.

As shown in an excerpt given in Table \ref{table:excerpt}, the columns of the Quasar Catalogue consist of the SDSS position (RA\textsubscript{q}, Dec\textsubscript{q}) and \textit{ugriz} magnitudes of 40,000 quasars. For the 23,779 of them for which pointing could be calculated, the calculated position (RA\textsubscript{p}, Dec\textsubscript{p})  of the optimal pointing are also provided, as is the score for that pointing.  For the other 16,221 quasars, no suitable pointing could be found.  A suitable pointing cannot be generated for a target for which no reference passes the filters shown in Section 2.2.  The reasons for this are elaborated in \citet{locuspaper}.  In these cases, the pointing coordinates and score are filled with a value of 0.

For each of the 23,779 quasars for which a pointing is presented, that pointing is the best possible pointing under the criteria specified in Section \ref{Parameters}.  Amongst these, the scores can be used to provide a ranking indicating the number and quality of reference stars for each target.  The distribution and relative meaning of these scores is discussed in Subsection \ref{Implications} below.

\subsection{Using the Catalogue}
\label{Using}
This Subsection describes the use of the catalogue by observers with observational constraints identical to those specified in Subsection \ref{Observing}.  For observers with observing constraints similar to, but not identical to these, the system can be used with certain caveats as outlined in Section \ref{Conclusion}.

The locus algorithm system is envisaged for use in two scenarios: one where an observer has a pre-determined target and one where the observer has a class of targets to choose from and uses the system to inform this choice.  In the first case, an observer can search through the catalogue for their pre-selected quasar, and the pointing listed there can be used to enable optimised photometry.

In the case of an observer who wishes to observe quasars more generally, the catalogue can be used to identify targets with many good reference stars and thus the best conditions for differential photometry.  Such observers can thus use the scores of the pointings for the targets provided to inform their choice of target (e.g. by selecting only those targets with a score in the 95\textsuperscript{th} percentile as shown in Table \ref{table:descriptives}). 

Having selected a target, the telescope should be aligned such that the edges of the FoV of its detector are aligned with the North/South and East/West axes, and the telescope pointed at the optimum pointing provided in the catalogue.  The optimum set of reference stars are not stored in the catalogue, but supplied with the catalogue is a pre-generated SQL query formatted suitably for SDSS CAS access which would enable the user to identify the individual reference stars to be used \citep{ZenodoQuasarCatalogue}.

\begin{table*}
\centering
\begin{tabular}{ |c|c||c|c|c|c|c||c|c|c| } 
\hline
 \multicolumn{7}{|c||}{\bfseries Quasar} & \multicolumn{3}{c|}{\bfseries Pointing} \\
\hline
 \multicolumn{2}{|c||}{\bfseries Position} & \multicolumn{5}{c||}{\bfseries Magnitude} & \multicolumn{3}{c|}{\bfseries Attributes} \\
\hline
RA\textsubscript{q} & Dec\textsubscript{q} & u & g & r & i & z & RA\textsubscript{p} & Dec\textsubscript{p} & Score \\ 
\hline
256.0224 & 59.0204 & 19.35 & 19.193 & 19.173 & 18.92 & 18.907 & 255.9825 & 58.9371 & 1.2341 \\ 
256.1418 & 59.201 & 19.373 & 19.261 & 19.035 & 18.964 & 19.086 & 256.1441 & 59.1177 & 2.3484 \\ 
255.5668 & 59.4488 & 19.537 & 19.411 & 19.085 & 19.053 & 19.034 & 255.6814 & 59.4744 & 1.8073 \\ 
255.4656 & 59.3727 & 19.073 & 18.84 & 18.73 & 18.716 & 18.77 & 255.6291 & 59.3082 & 0.7637 \\ 
\rowcolor{pink}
255.5836 & 59.2607 & 20.266 & 19.816 & 19.314 & 18.778 & 18.482 & 0 & 0 & 0 \\ 
\rowcolor{pink}
255.2419 & 60.3599 & 24.579 & 21.522 & 19.838 & 19.696 & 19.362 & 0 & 0 & 0 \\ 
255.2369 & 60.4444 & 19.766 & 19.138 & 18.783 & 18.554 & 18.241 & 255.3836 & 60.4569 & 1.8138 \\ 
255.0467 & 60.0616 & 19.747 & 19.287 & 18.672 & 18.072 & 17.704 & 255.2137 & 60.001 & 0.9387 \\ 
256.0256 & 60.7983 & 18.56 & 18.643 & 18.292 & 18.201 & 18.191 & 255.9964 & 60.7384 & 4.103 \\ 
\rowcolor{lime}
255.9825 & 60.7533 & 19.601 & 19.222 & 18.766 & 18.616 & 18.283 & 255.8417 & 60.7211 & 5.2643 \\ 

 \hline
 \end{tabular} 
 \caption{Excerpt from Quasar Catalogue \citep{ZenodoQuasarCatalogue}.  Columns are Right Ascension, Declination, Magnitude (u, g, r, i, z), Pointing RA, Pointing Dec and Score.  Highlighted in red are two quasars for which no suitable pointings were possible for the given criteria.  Highlighted in green is the quasar with the best score in this small sample, SDSS J170355.79+604511.7  Excerpt from \citet{ZenodoQuasarCatalogue}, layout copied from \citet{creaner2016thesis}}
 \label{table:excerpt} 
\end{table*}

\subsection{Implications of Scoring}
\label{Implications}
\begin{figure}
        \center{\includegraphics[width=0.47\textwidth]
        {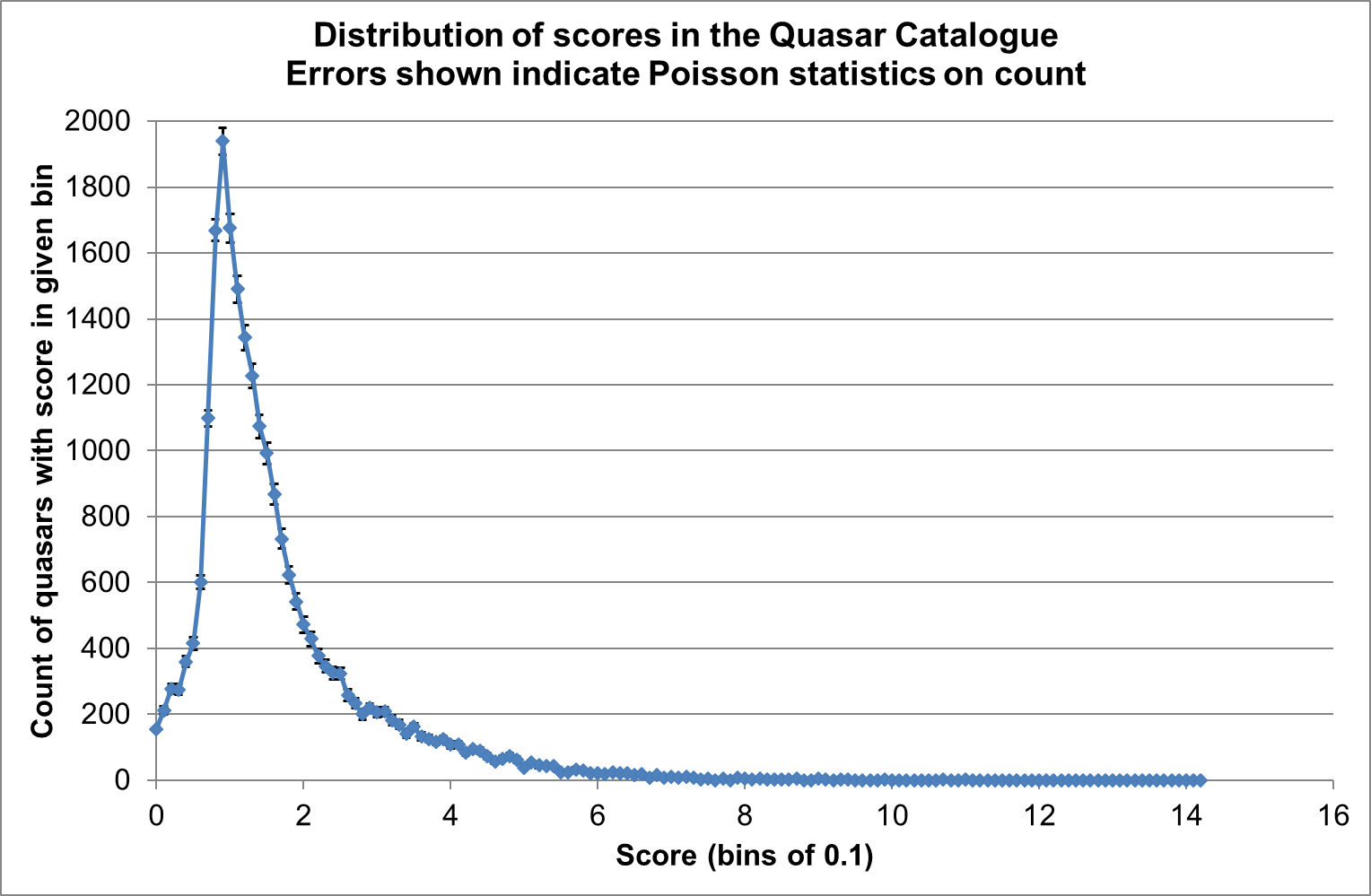}}
        \caption{\label{fig:quasar_count}  Distribution of scores in the Quasar Catalogue.  This graph shows only the distribution of scores for 23,779 quasars for which a valid pointing could be identified.  16,221 quasars for which no pointing was generated are not shown.  A bin size of 0.1 points of score is used to show this distribution smoothly.  Errors shown on the graph indicate the Poisson Statistic of the count. As can be seen, most quasars have a relatively low score. There are some quasars with a relatively high score.  These are quasars whichmultiple suitable references nearby.  }
      \end{figure}

This Subsection provides a set of statistical measures of the scores as shown in Table \ref{table:descriptives} and Figure \ref{fig:quasar_count} shows the distribution of scores for quasars in the catalogue. This is intended to allow an observer to interpret the score for a given target relative to the scores of other targets in the catalogue.   The scores presented in figure 2 are calculated using the \texttt{best\textunderscore{}mag} parameters from SDSS.Using SDSS magnitude errors, the variation in score is less than 1\% for targets of magnitude 18 and brighter.

Table \ref{table:descriptives} provides a table of descriptive statistics of the data within the catalogue. These describe the range and distribution of the values in the catalogue in summary form and allow an observer to determine broadly where within that distribution a given target lies.

An observer using this catalogue to choose targets based upon their suitability for differential photometry may determine a suitable cut-off score for their observing conditions, or may select only those targets with the highest scores upon which to focus their observations using the methods described at \citet{ZenodoQuasarCatalogue}. 

An observer who has already determined their target may search this catalogue using the data provided at \citet{ZenodoQuasarCatalogue} to determine whether the target is among the 23,779 targets for which a pointing has already been generated, and thus use that pointing.  These users may also wish to further optimise the pointing for their target by running the software at \citet{githubrepo} with the observing parameters tuned for their own equipment and location.

\subsection{Score Distribution}
\label{distribution}
As can be seen from Figure \ref{fig:quasar_count}, the distribution of scores has a sharp peak and a long, right tail.  What this means is that most quasars have a relatively low score, but that there are some quasars with a relatively high score, i.e. quasars which serendipitously have several suitable references nearby. From Table \ref{table:descriptives} it can, for example, be seen that the 95\textsuperscript{th} percentile for scores is 4.2493.  This means that 1,184 quasars have pointings with scores greater than that under this set of observing parameters.  
      
The peak of the distribution of the scores is
$\le$
2.  Low scores indicate quasars for which there are few references with similar colour and magnitude to the quasar in question. This is not unexpected since, as \citet{fan1999simulation} states, one method for differentiating quasars from stars is the fact that quasars show a generally different distribution of colour index when compared with stars.  A close match between target and reference on this very colour index is essential to a high rating for a reference, and multiple high-rating references are required for a high score.  Because of this disparity, it follows that for many quasars, few or no suitable reference stars can be identified.

\begin{figure}
        \center{\includegraphics[width=0.47\textwidth]
        {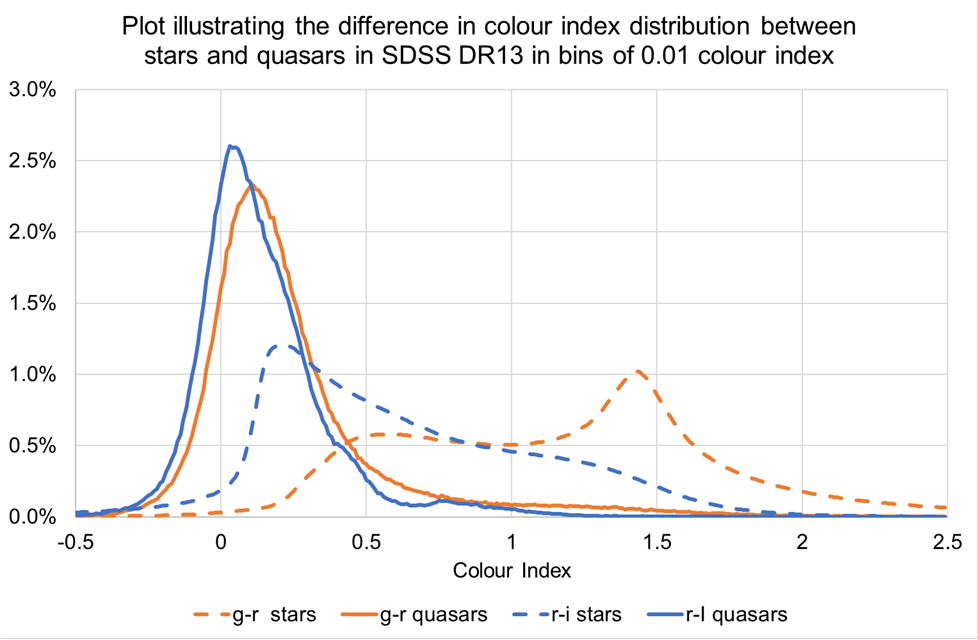}}
        \caption{\label{fig:qso_star_colours}  Colour index distribution for stars and quasars in SDSS-\textit{g-r} and \textit{r-i}s.  Data extracted from SDSS DR13 \citep{2017ApJS..233...25A}}  
      \end{figure}

The peak of the distribution of the scores is
$\le$
 2. Low scores
indicate quasars for which there are few references with similar colour and magnitude to the quasar in question. This is not unexpected since, as \citet{fan1999simulation} states, one method for differentiating quasars from stars is the fact that they have intrinsically different spectral properties. This is borne out by Figure \ref{fig:qso_star_colours} that shows the difference in colour index distribution between stars and quasars. In this figure, the solid lines depict the \textit{g-r} (orange) and \textit{r-i} (blue) colour differences for quasars, the dashed lines represent the same for stars. The disparity betewen these lines is testimony to the difference in colour properties of stars and quasars.

This disparity is further born out when the descriptive statistics of the quasar catalogue are examined.  90\% (5\textsuperscript{th} to 95\textsuperscript{th} percentile) of identified quasars in SDSS are found between the SDSS \textit{g}-\textit{r} colour indices -0.0170 and 0.5981, and \textit{r}-\textit{i} of -0.0580 and 0.3450 \citep{creaner2016thesis}. By contrast, the corresponding values for the 5\textsuperscript{th} to 95\textsuperscript{th} percentile for colour indices for stars in SDSS are \textit{g}-\textit{r} of 0.27 to 1.86 and an \textit{r}-\textit{i} of 0.05 to 1.63\citep{creaner2016thesis}.  Since stars are used as the references, it follows that the small overlap in the colour distribution between the star and quasar populations implies relatively few available references to use in determining an ideal pointing. 

It should be noted that even in the case of quasars with low scores, the pointing that is presented in this catalogue is the best available pointing for that quasar under the specified observational conditions.

\begin{table*}
\centering
\begin{tabular}{ |c||c||c|c||c|c||c| } 
\hline
 \multicolumn{1}{|c||}{\bfseries Descriptive} 
 & \multicolumn{3}{c||}{\bfseries Magnitude}  
 & \multicolumn{2}{c||}{\bfseries Colour}  
 & \multicolumn{1}{c|}{\bfseries Pointing} \\
\hline
{\bfseries Statistic} & {\bfseries g} & {\bfseries r} & {\bfseries i} &
{\bfseries g-r}& {\bfseries r-i} & {\bfseries score}  \\
\hline
{\bfseries Maximum} & 24.431 & 22.316 & 21.835 & 2.828 & 1.691 & 14.139 \\ 
{\bfseries 95\textsuperscript{th} Percentile} & 20.613 & 20.347 & 20.161 & 0.5981 & 0.345 & 4.249 \\ 
{\bfseries Mean} & 19.461 & 19.226 & 19.095 & 0.235 & 0.131 & 1.707 \\ 
{\bfseries Median} & 19.392 & 19.174 & 19.047 & 0.204 & 0.119 & 1.331 \\ 
{\bfseries 5\textsuperscript{th} Percentile} & 18.245 & 18.035 & 17.937 & -0.017 & -0.058 & 0.475 \\ 
{\bfseries Minimum} & 15.432 & 15.244 & 15.184 & -0.424 & -0.302 & 0.0008 \\ 
{\bfseries Standard Deviation} & 0.767 & 0.720 & 0.692 & 0.225 & 0.139 & 1.222 \\ 
 \hline
 \end{tabular} 
 \caption{Descriptive statistics of the variables in Quasar Catalogue, filtered to those quasars for which pointings were available.  Only the magnitude (g, r, i) and colour parameters (g-r, r-i) which contribute to the score are shown.    Copied from \citet{creaner2016thesis} Each column of descriptive statistics refers to the statistical value for that column only (e.g. the maximum r magnitude refers to the highest r magnitude of any target.  The maximum g value does not necessarily refer to the same target, but rather the target with the maximum g)}
 \label{table:descriptives} 
\end{table*}

\section{Discussion and Conclusions}
\label{Conclusion}

A catalogue has been developed to enable optimum photometry for a set of 23,779 quasars in the SDSS footprint, based on a particular set of observational criteria.  This catalogue, available from Zenodo at \citet{ZenodoQuasarCatalogue} can be used directly for observers whose observational criteria are identical to those used to generate it.  

This catalogue demonstrates the successful use of the system first proposed in \citet{creaner2010large} and described in \citet{locuspaper}.  Optimum pointings have been generated for photometry for each of 23,779 quasars in the 4th SDSS quasar catalogue.  The distribution of scores for these quasar pointings indicates that, for many quasars, few reference stars are available with close colour matches, but it is possible to identify a small number of quasars with highly suitable references by selecting quasars with high scores.  

An observer with observing conditions identical to those discussed in this paper may use this catalogue to identify targets upon which optimal differential photometry may be carried out, assuming no other \textit{a priori} conditions exist.  In addition, an observer may select priority targets amongst a list of candidates which they have produced by selecting those of their candidates for which the score is highest, indicating the highest potential for good observation.

An observer may use the instructions that accompany the catalogue on Zenodo in \citet{ZenodoQuasarCatalogue} to determine the identities of the reference stars for a given pointing. They may thus refine their choice of references or targets for their particular application based upon that output. 

\subsection{Target Selection}
\begin{figure}
        \center{\includegraphics[width=0.47\textwidth]
        {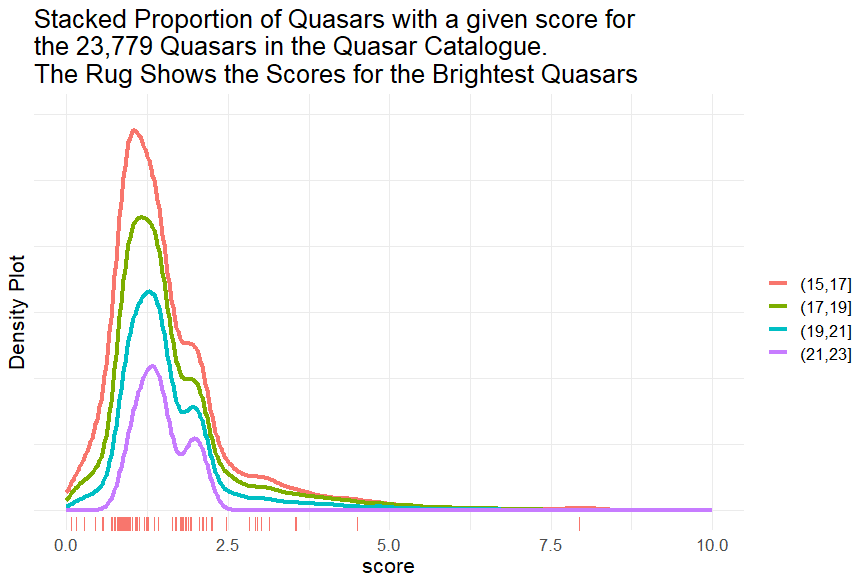}}
        \caption{\label{fig:QSO_score_distribution} Variation of the distribution of scores by quasar \textit{r}-magnitude.  This graph shows only the distribution of scores for 23,779 quasars for which a valid pointing could be identified.  16,221 quasars for which no pointing was generated are not shown. }
      \end{figure}

The distribution of scores with respect to magnitude shown in Figure \ref{fig:QSO_score_distribution} approximately matches to the distribution of quasars in the 4th SDSS catalogue. It shows that there is only a weak dependence of score on Quasar magnitude. The figure therefore shows that there reasonable scores across all magnitude ranges for quasars in the catalogue. As a result, an observer wising to select a quasar may select one within their observational capability so long as their system is sensitive to the r-magnitude range 18.035 to 20.347 in which 90\% of the quasars in the catalogue are found as shown in Table \ref{table:descriptives}.

\subsection{Changing observing conditions}
\begin{figure}
        \center{\includegraphics[width=0.47\textwidth]
        {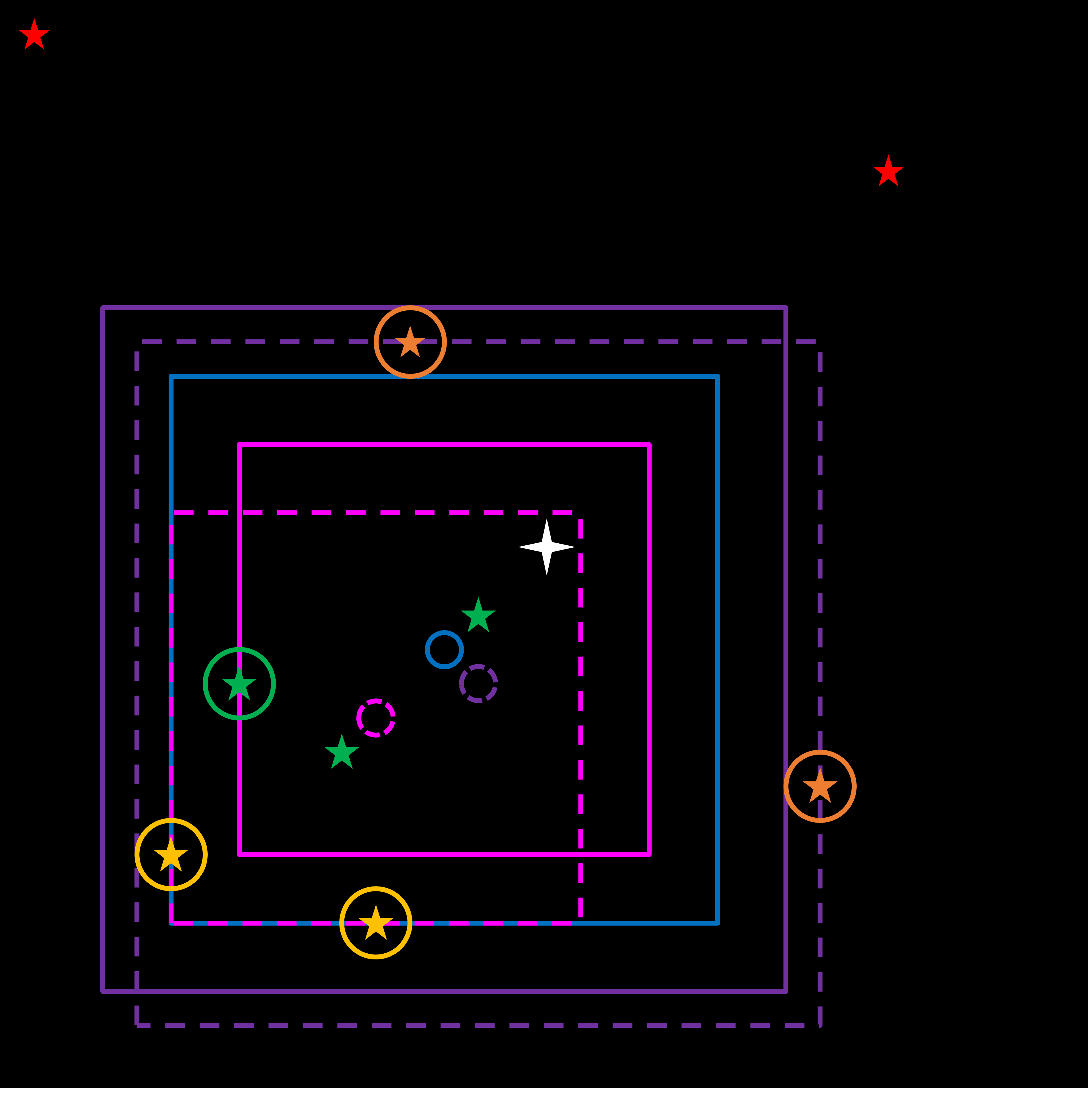}}
        \caption{\label{fig:locus_alternate} Locus Algorithm with different FoV sizes.  Target, references and pointing identical to Figure \ref{fig:locus_figure}.  Blue outline shows the original pointing.  Purple and Magenta show larger and smaller FoV sizes respectively.  Solid lines show results of using the originally calculated pointing.  Dashed lines show the optimal pointings for those FoV sizes.  Circled stars indicate those references affected by the size of FoV. The red stars are not included in the FoV at any size.  The green stars are included at all three sizes.  The stars in yellow are excluded from the smaller FoV if the original pointing is used, but can be included with the modified pointing (dashed magenta).  The orange stars are not included in the FoV with the original pointing, but can be included with the larger FoV.   Modified from \citet{creaner2016thesis}}
      \end{figure}

An observer with similar, but not identical conditions, can use this catalogue subject to the following considerations.  Firstly, an observer looking to identify targets with many similar reference stars choose use high-scoring targets in this catalogue to observe.  However, if their FoV size is different to that used to generate this catalogue, there are some limitations as illustrated in Figure \ref{fig:locus_alternate}.

The Locus algorithm, by definition, places at least two stars at the edge of the usable field of view.  This means that observations made with a narrower FoV than that for which the catalogue was optimised automatically lose two, and possibly more, designated reference stars, as shown in Figure \ref{fig:locus_alternate}, with the magenta squares.  As a result, the photometry for the smaller field may be compromised, and therefore pointings optimised for a FoV larger than that of the telescope at the observatory should not be used. 

On the other hand, if an observatory with a larger FoV uses the output of a catalogue optimised for a smaller FoV, the FoV will automatically include all of the reference stars included in the smaller one.  However, the larger FoV may include stars which would have been viable references, but which have been excluded from consideration by the algorithm for the smaller FoV.  In addition, it may have been possible to reposition the larger field to include even more references as illustrated with the purple squares in Figure \ref{fig:locus_alternate}. 

It may therefore be concluded that in the absence of a perfectly optimised catalogue, the pointings optimised for a smaller FoV may be used, and will work at least as well on a telescope with a larger one, though they will not be optimised for that telescope.

\subsection{Further Work with the Locus Algorithm}
Scoring is a modular subsystem of the code, which can be replaced with new systems to suit the requirements of an observer. Because the scoring system is based on parameters defined by the user at run-time, it is possible to reuse the algorithm with different parameters and this will lead to different pointings being generated for the same target, with different references and different scores. 

The highly flexible, extensible and modular system developed to create this catalogue allows for further catalogues to be generated with different observational criteria (target sets, FoV size, limiting magnitude and colour etc) using the Locus Algorithm. This system can and is intended to be reused and refined to provide further catalogues of pointings under different observing conditions, and to provide pointings for other classes of object, such as the Exoplanet Catalogue presented in \citet{ZenodoXOPCatalogue}.  

\section*{Acknowledgements}
\textbf{Funding for this work}: This publication has received funding from Higher Education Authority Technological Sector Research Fund and the Institute of Technology, Tallaght, Dublin Continuation Fund (now Tallaght Campus, Technological University Dublin).

\textbf{SDSS Acknowledgement}: This paper makes use of data from the Sloan Digital Sky Survey (SDSS).  Funding for the SDSS and SDSS-II has been provided by the Alfred P. Sloan Foundation, the Participating Institutions, the National Science Foundation, the U.S. Department of Energy, the National Aeronautics and Space Administration, the Japanese Monbukagakusho, the Max Planck Society, and the Higher Education Funding Council for England. The SDSS Web Site is http://www.sdss.org/.

The SDSS is managed by the Astrophysical Research Consortium for the Participating Institutions. The Participating Institutions are the American Museum of Natural History, Astrophysical Institute Potsdam, University of Basel, University of Cambridge, Case Western Reserve University, University of Chicago, Drexel University, Fermilab, the Institute for Advanced Study, the Japan Participation Group, Johns Hopkins University, the Joint Institute for Nuclear Astrophysics, the Kavli Institute for Particle Astrophysics and Cosmology, the Korean Scientist Group, the Chinese Academy of Sciences (LAMOST), Los Alamos National Laboratory, the Max-Planck-Institute for Astronomy (MPIA), the Max-Planck-Institute for Astrophysics (MPA), New Mexico State University, Ohio State University, University of Pittsburgh, University of Portsmouth, Princeton University, the United States Naval Observatory, and the University of Washington.

This paper makes use of \textbf{CFITSIO}: A FITS File Subroutine Library by \citet{pence1999cfitsio}




\bibliographystyle{mnras}
\bibliography{quasarpaper}




%
%


\bsp	
\label{lastpage}
\end{document}